\documentclass[prl,twocolumn,showpacs]{revtex4}

\newcommand{\Eq}[1]{Eq.~(\ref{#1})}
\newcommand{\Eqs}[1]{Eqs.~(\ref{#1})}
\newcommand{\Fig}[1]{Fig.~\ref{#1}}

\newcommand{\teq}{\tau_\mathrm{eq}}
\newcommand{\tmax}{\tau_\mathrm{max}}
\newcommand{\taver}{\tau_\mathrm{aver}}
\newcommand{\Tmin}{T_\mathrm{min}}
\newcommand{\Tmax}{T_\mathrm{max}}
\newcommand{\Uperp}{\Upsilon_\perp}
\newcommand{\Upara}{\Upsilon_\parallel}
\newcommand{\Jpara}{J_\parallel}

\newcommand{\Irms}{I_\mathrm{rms}}
\newcommand{\av}[1]{[#1]_\mathrm{av}}

\usepackage{amsmath}
\usepackage{graphicx}

\begin{document}
\title{Vortex glass transition in a frustrated 3D XY model with disorder} 

\author{Peter Olsson}

\affiliation{Department of Physics, Ume\aa\ University, 
  901 87 Ume{\aa}, Sweden}

\date{\today}   

\begin{abstract}
  The anisotropic frustrated three dimensional (3D) XY model with
  disorder in the coupling constants is simulated as a model of a
  point disordered superconductor in an applied magnetic field.  From
  a finite size scaling analysis of the helicity modulus it is
  concluded that the data is consistent with a finite temperature
  transition with isotropic scaling and the correlation length
  exponent is found to be $\nu=1.50\pm0.12$, consistent with 3D gauge
  glass universality.
\end{abstract}


\maketitle

The hypothesis of a vortex glass in disordered high temperature
superconductors\cite{Fisher:89,Fisher_Fisher_Huse} has spurred much
research and many discussions during more than one decade and
continues to be a very controversial
issue\cite{Reichhardt_Otterlo_Zimanyi}.  The essence of this
suggestion is that random point disorder in superconductors may
conspire with the vortex line interaction to pin the vortices and that
this takes place through a sharp transition into a phase with
vanishing linear resistance.

Computer simulations have for quite some time played an important role
in the examination of critical phenomena and have also recently been
used in the study of some vortex glass models.  One important such
model is the three dimensional (3D) gauge glass, which is an isotropic
3D XY model with randomness included through a random vector potential
added to the phase difference of the superconducting order parameter.
The evidence has for quite some time pointed at a finite temperature
transition in this model\cite{Huse_Seung,Reger_TYF}, but strong
evidence for a real phase transition has been obtained only
recently\cite{Olson_Young} through the use of the exchange Monte Carlo
(MC) technique\cite{Hukushima_Nemoto}. The value of the correlation
length exponent was then found to be $\nu = 1.39\pm0.20$.  It is
however generally recognized that the 3D gauge glass models is too
much of a simplification to allow for any safe conclusions regarding
the behavior of disordered superconductors in applied magnetic
fields\cite{Huse_Seung}.  Most seriously, the gauge glass is an
isotropic model with no net field which means that the possibility of
anisotropic scaling is excluded at the outset.

Two studies with the necessary ingredients of disorder and applied
field have so far been reported in the literature. The first is an
examination of a frustrated 3D XY model with randomness in the
couplings\cite{Kawamura:00}, and the data was there interpreted as
evidence for a phase transition with $\nu\approx 2.2$.  The crossing
of the data for different sizes expected from finite size scaling, was
however not entirely convincing, possibly because of the open boundary
conditions employed in the simulation. The second study gives results
for a 3D random pinning model with strong
disorder\cite{Vestergren_Lidmar_Wallin}. The value of the correlation
length exponent was there found to be $\nu\approx 0.7$,
undistinguishable from $\nu\approx 0.67$ in the pure zero field 3D XY
model. This is at odds with the common expectation that a vortex glass
transition should be in a different universality class than the pure
model.

In this paper we present results from large scale simulations on a
frustrated 3D XY model with disorder in the coupling constants. The
quantity in focus is the helicity modulus and we find that the data is
consistent with a finite temperature glass transition with isotropic
scaling and obtain the correlation length exponent $\nu=1.50\pm0.12$.
The agreement with $\nu=1.39\pm0.20$ for the 3D gauge glass
model\cite{Olson_Young} suggests a common universality class.

The model we simulate is given by the
Hamiltonian\cite{Olsson_Teitel:xy3fp}
\begin{equation}
  {\cal H} = -\sum_{{\rm bonds}\,i\mu}J_{i\mu}\cos
  (\theta_i-\theta_{i+\hat\mu}-A_{i\mu} + \delta_\mu),
  \label{Hamiltonian}
\end{equation}
where $\theta_i$ is the phase of the superconducting wave function at
site $i$ of a periodic $L_x\times L_y\times L_z$ lattice and the sum
is over all bonds in directions $\mu = x$, $y$, $z$.  An applied
magnetic field in the $z$ direction corresponding to $1/5$ flux
quantum per plaquette is obtained through the quenched vector
potential with the choice $A_{ix} = y_i\; 2\pi/5$, and $A_{iy} =
A_{iz} = 0$. The randomness is included through disorder in the
coupling constants,
\begin{displaymath}
  \begin{array}{ll}
    J_{i\mu}=J_\perp(1+p\epsilon_{i\mu}),\qquad & \mu=x,y, \\
    J_{i\mu}=\Jpara, & \mu=z,
  \end{array}
\end{displaymath}
where $\epsilon_{i\mu}$ are independent variables from a Gaussian
distribution with $\langle\epsilon_{i\mu}\rangle = 0$ and
$\langle\epsilon_{i\mu}\rangle^2 = 1$. The disorder strength $p=0.4$
together with the anisotropy $\Jpara = J_\perp/40$ were choosen since
they were found sufficient to prohibit the formation of Abrikosov
lattices.  The simulations are performed with fluctuating twist
boundary conditions\cite{Olsson:self-cons.long}; the $\delta_\mu$ in
\Eq{Hamiltonian} are the twist variables and the total twist in the
respective directions are $\Delta_\mu = L_\mu\delta_\mu$.  The
simulations are performed with $L = L_x = L_y$ and a fixed aspect
ratio, $L/L_z = 5/3$. The temperature is given in units of $J_\perp$.

The quantity in focus in our analysis is the helicity modulus which is
defined from the change in free energy density, $f$, due to an applied
twist, $\delta_\mu$: $\Upsilon_\mu = \partial^2 f/\partial
\delta_\mu^2$ \cite{Fisher_Barber_Jasnow}. To use the helicity modulus
as a signal of the stiffness of the system the derivative should be
evaluated at the twist that minimizes the free energy.  In ordered
systems this minimum is always at zero twist and the helicity modulus
may then be evaluated by means of a correlation function determined
with periodic boundary conditions, $\Delta_\mu=0$. For disordered
systems, however, the minimizing twist will in general be different
from zero and 
one then needs to make simulations with the twist variables
$\Delta_\mu$ as additional dynamical variables and collect histograms
$P_\mu(\Delta_\mu)$.  The helicity modulii are then determined from
the free energies $F_\mu = -T\ln P_\mu$, as discussed below. To
analyze the critical behavior we use the standard scaling relation for
the helicity modulus in 3D,
\begin{equation}
  L\Upsilon \sim g\left(t L^{1/\nu}\right),
  \label{scaling}
\end{equation}
where $t=(T-T_c)/T_c$ and $g$ is a scaling function\cite{Li_Teitel:89}. This expression
presumes isotropic scaling; the more general scaling relations are
gives in Ref.\ \cite{Vestergren_Lidmar_Wallin}. A naive analysis of
$\Upsilon(L)$ rather than the correct scaling quantity $L\Upsilon$,
led to an erroneous conclusion regarding the existence of a vortex
glass phase in Ref.\ \cite{Olsson_Teitel:xy3fp}.
\begin{table}[htb]
  \begin{tabular}{|c|c|c|c|l|c|c|c|c|}
   $L$ & $L_z$ & $N_d$ & $N_T$ & $\Tmin$ & $X_\mathrm{acc}$ (\%) &
   $\teq$ & $\tmax$ & sweeps$/10^6$\\
   \hline
   $10$ & $6$  & 600 & 12 & 0.09  & 30 &  1 & 15 &  0.3 + 3.9 \\
   $15$ & $9$  & 600 & 24 & 0.09  & 30 &  4 & 12 &  1.0 + 3.1 \\
   $20$ & $12$ & 600 & 36 & 0.09  & 32 & 11 & 21 &  2.9 + 5.5 \\
   $25$ & $15$ & 200 & 36 & 0.115 & 27 & 17 & 31 &  4.5 + 8.1 
  \end{tabular}
  \caption{Parameters describing the simulations. For systems of size
   $L\times L\times L_z$ we simulated $N_d$ disorder configurations
   with $N_T$ temperatures in the range $\Tmin \leq T < \Tmax$, cf.\
   \Eq{Tm}. The acceptance ratio for the exchange step is 
   given by $X_\mathrm{acc}$. Of the bins corresponding to
   $2^{18}=262144$ sweeps $\teq$ are first discarded and the remaining
   $\tmax$ are used for calculating averages. The same information is
   also given in terms of the number of sweeps for equilibration and
   for collecting data.} 
  \label{Param}
\end{table}

Our simulations are performed with exchange MC which is a method for
simultaneously simulating multiple copies of a particular
configuration of disorder with each copy at a different temperature.
According to certain rules these copies may now and then interchange
temperature\cite{Hukushima_Nemoto} and therefore effectively perform
random walks in temperature space. These random changes in temperature
greatly help the different copies avoid getting trapped in restricted
parts of the phase space and therefore makes it possible to sample
the whole phase space and obtain the true thermodynamic averages. In
our simulations the temperatures were chosen according to the equation
\begin{equation}
  T_m = \Tmin \left(\frac{\Tmax}{\Tmin}\right)^{m/N_T},\quad
  m=0,\ldots,N_T-1,
  \label{Tm}
\end{equation}
with $\Tmax = 0.24$ and $\Tmin$ as given in Table \ref{Param}.  Before
doing the actual exchange MC the initial spin configurations for the
$N_T$ temperatures were obtained by slowly cooling the system with
standard MC simulations. The number of temperatures, the acceptance
ratio for the exchange step, and the number of disorders simulated for
the different system sizes are shown in Table \ref{Param}.  The
exchange steps are attempted once every 16 sweep.

The exchange MC method ensures that all the different copies remain at
thermal equilibrium as soon as equilibrium has been reached.  The
approach to equilibrium may however be very slow since information and
configurations have to propagate all the way from high to low
temperatures.  We have carefully examined the approach to equilibrium
and especially for our largest sizes the times for equilibration are
indeed very long.  For the next largest size, $L=20$, equilibration is
only reached after about $2.9\times10^6$ sweeps, and to make
thermalization at all possible for our largest size, $L=25$, we chose
not to go to quite that low temperatures for the largest size, cf.\ 
Table \ref{Param}. This was decided since the time required for
thermalization may increase very rapidly with decreasing temperature.
Still, about $4.5\times 10^6$ sweeps were necessary to reach
equilibrium for $L=25$.

\begin{figure}[htbp]
  \includegraphics[width=8cm]{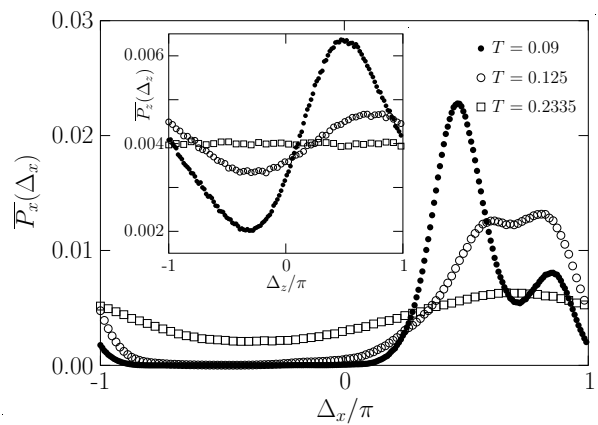}
  \caption{$\overline{P_x}(\Delta_x)$ for a certain disorder
    realisation. Note that the peaks of the histograms become higher
    and sharper as the temperature is lowered.  This kind of data is
    used for the determination of $\Upsilon_x$ through \Eq{Upsilon}.
    The inset gives the corresponding quantity for the $z$ direction.}
  \label{PDelta}
\end{figure}

For each system size, disorder configuration and temperature the main
output from the simulations are histograms $P_\mu(\Delta_\mu; \tau)$,
where $\tau$ enumerates bins corresponding to $2^{18} = 262144$ sweeps
over the lattice.  The further analysis is then based on the average
\begin{equation}
  \overline{P_\mu}(\Delta_\mu) = \frac{1}{\tmax}
  \sum_{\tau=1}^{\tmax} P_\mu(\Delta_\mu; \tau),
  \label{overlineP}
\end{equation}
and the helicity modulus is determined from the curvature at the
minimum of the associated free energy by fitting a second order
polynomial to the free energy in a narrow interval around the minimum,
$\Delta_\mu^0$, which is also determined in the fit,
\begin{equation}
  \overline{F_\mu}(\Delta_\mu) \sim \Upsilon_\mu \frac{V}{2L_\mu^2}
  (\Delta_\mu - \Delta_\mu^0)^2.
  \label{Upsilon}
\end{equation}
However, it turns out that the values of $\Upsilon_\mu$ obtained in
this way are biased towards too large values. The origin of this bias
as well as the method employed to eliminate it from the data is
discussed below after the discussion of the results.

\begin{figure}[htbp]
  \includegraphics[width=8cm]{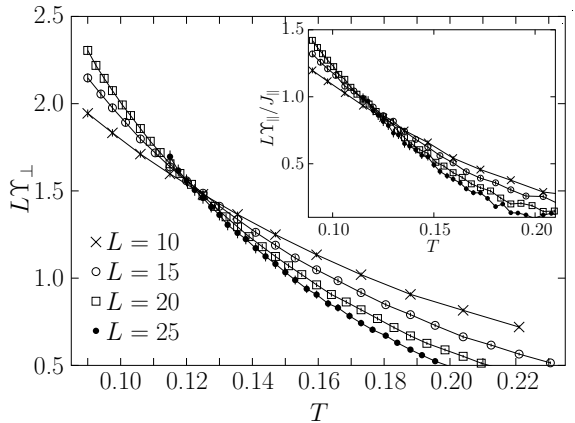}
  \caption{$L\Uperp$ versus temperature for four different system
    sizes. The curves cross at a single point, which is an indication
    of critical behavior. The inset shows $L\Upara/\Jpara$ which shows
    a similar crossing at almost the same temperature.}
  \label{Ups}
\end{figure}

\begin{figure}[htbp]
  \includegraphics[width=8cm]{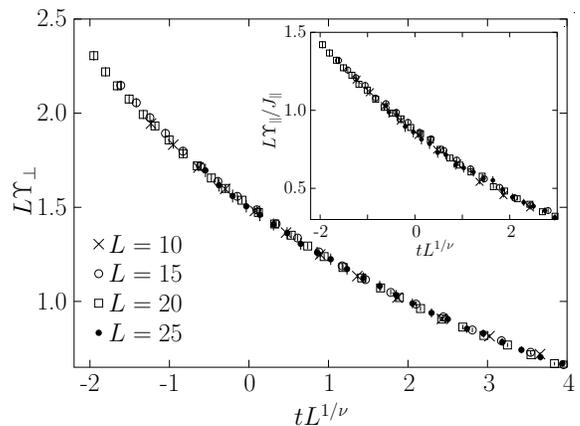}
  \caption{The scaling collapse of $L\Uperp$ gives $\nu = 1.50\pm0.12$
    and $T_c = 0.123\pm 0.003$. The inset is an attempt to collapse
    $L\Upara/\Jpara$ with the same values for $\nu$ and $T_c$. The
    nice collapse gives additional support for a vortex glass
    transition with isotropic scaling.}
  \label{Ucoll}
\end{figure}

We now focus on disorder averaged quantities for which the bias
mentioned above has already been eliminated. With $\av{\ldots}$
denoting the average over disorder configurations we define
\begin{subequations}
  \label{DisAver}
\begin{eqnarray}
  \Uperp & = & \frac{1}{2}\av{\Upsilon_x + \Upsilon_y}, \\
  \Upara & = & \av{\Upsilon_z}.
\end{eqnarray}
\end{subequations}
Our results for $L\Uperp$ are shown in \Fig{Ups}. To a very good
accuracy the data for the different sizes cross at a single
temperature. To further verify the scaling according to \Eq{scaling}
we fit our data for $L\Upsilon_\perp$ near $T_c$ to a fourth order
polynomial expansion of $g(tL^{1/\nu})$ and obtain the values $\nu =
1.50\pm0.12$ and $T_c = 0.123\pm 0.003$. The collapse which is shown in
\Fig{Ucoll} is excellent and holds in a surprisingly large temperature
interval. The error estimates are obtained with a resampling technique
and correspond to one standard deviation.

A vortex glass transition should also be seen in the parallel
component of the helicity modulus and since the anisotropy exponent
enters the scaling relation in different ways for $\Upara$ and
$\Uperp$\cite{Vestergren_Lidmar_Wallin}, a scaling analysis of
$L\Upara$ constitutes an additional test that the scaling actually is
isotropic.  The crossing of $L\Upara$ at $T_c$ for different sizes and
the scaling collapse are shown in the insets of Figs.~\ref{Ups} and
\ref{Ucoll}, respectively.  From the not so smooth curves it is clear
that the precision in $\Upara$ is much worse than for $\Uperp$.  This
may be traced back to the less good quality of the histograms
$P_z(\Delta_z)$, as shown in the inset of \Fig{PDelta}.  Nevertheless,
the data for $L\Upara$ (see inset in \Fig{Ucoll}) collapses nicely
when using the same values of $\nu$ and $T_c$ as in the main figure.
This therefore constitutes additional evidence for a vortex glass
transition in the model.  We now turn to the bias mentioned above and
also discuss similarities and differences with the more common method
to determine the root mean square current $\Irms$, in analyses of
models with disorder.

As mentioned above the determination of $\Upsilon$ from a twist
histogram suffers from a bias towards too large values. To investigate
the reason for this bias we performed additional simulations with
fluctuating twist boundary conditions of the pure zero-field 3D XY
model and collected histgrams $P(\Delta,\tau)$ where $\tau$ enumerates
the bins. This data was then used to calculate averages over $\taver$
consecutive bins $\overline{P}(\Delta;\taver)$, which in turn were
used to determine the helicity modulus. By repeating this procedure
for several values of $\taver$ the dependence of $\Upsilon$ on
$\taver$ was determined. From this kind of analysis it was found that
the bias decays as $1/\taver$ to a very good precision. Another
finding is that the bias may be made to vanish altogether by making
use of $\Delta^0 = 0$ in \Eq{Upsilon} instead of using $\Delta^0$ as a
free parameter. ($\Delta^0 = 0$ is the known value of the minimizing
twist in the pure system.) The latter observation suggests that the
bias is related to the usual complication in determining the width
(variance) of a distribution when the true average is not
known\cite{Variance}. Since $\Upsilon$ is inversely related to the
variance of the distribution $P(\Delta)$ around the maximum it follows
that the estimates based on runs of length $\taver$ would be expected
to decay towards the true value as\cite{Variance}
\begin{equation}
  \Upsilon(\taver) = \frac{1}{1-b/\taver} \Upsilon(\infty),
  \label{Ut}
\end{equation}
where $b$ is a free parameter related to the decorrelation time in the
simulations. For small values of $b/\taver$ the expected behavior is
therefore entirely in accordance with the $1/\taver$ decay discussed above.

\begin{figure}[htbp]
  \includegraphics[width=8.5cm]{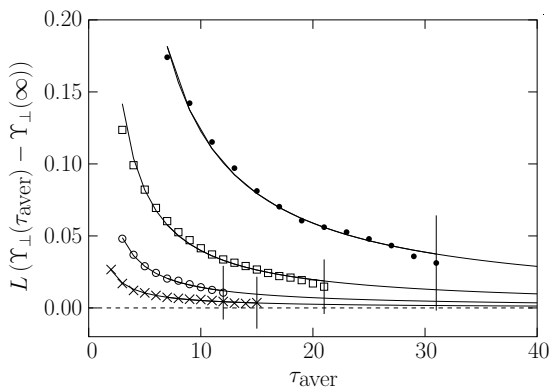}
  \caption{The figure shows the elimination of the bias in $\Uperp$ by
    fitting $\Uperp(\taver)$ to \Eq{Ut} with $b$ and $\Uperp(\infty)$
    as free parameters. The time is in units of the bin size which is
    $2^{18} = 262144$ sweeps.}
  \label{Uperp-tau}
\end{figure}

Returning to our data for the vortex glass model, the procedure used
to determine the data points in \Fig{Ups} consists of three steps: (i)
Determine $\Upsilon_\mu(\taver)$ for each disorder configuration and
several values of $\taver$ by fitting histogram
$\overline{P}_\mu(\Delta_\mu;\taver)$ based on $\taver$ consecutive
bins, $P_\mu(\Delta_\mu,\tau)$ to \Eq{Upsilon}. (ii) Calculate the
disorder averaged quantities $\Uperp(\taver)$ and $\Upara(\taver)$,
cf.\ \Eqs{DisAver}.  (iii) Fit this data to \Eq{Ut} to obtain the
unbiased estimates $\Uperp \equiv \Uperp(\infty)$ and $\Upara \equiv
\Upara(\infty)$.  The last step is illustrated in \Fig{Uperp-tau} for
$T=0.125$ close to $T_c$.  The error bars on the last point for each
size are the errors associated with the disorder average.

We now shortly discuss the relation between the helicity modulus
calculated in the present paper and the more common method to
determine the root mean square current $\Irms$\cite{Reger_TYF}.
Since the current and the helicity modulus are first and second
derivatives, respectively, of the same function $F(\Delta)$, both
quantities effectively probe the roughness of the function
$F(\Delta)$.  The low temperature phase is characterized by large
energy barriers growing with system size whereas the free energy above
the transition temperature becomes a flat function of $\Delta$ in the
limit of large $L$.  However, $\Upsilon$ turns out to be more
efficient in measuring the roughness of $F(\Delta)$. The reason is
that $\Upsilon$ measures a property at an extremum (the minimum of the
free energy) whereas the current at $\Delta=0$ for a given shape of
the function may be large or small depending on the location of the
structure in $F(\Delta)$.  $\Irms$ is therefore only \emph{on the
  average} a good measure of the properties of $F(\Delta)$ and this
has to be compensated for by using a larger number of disorder
configurations.  This is the reason for the good precision in our data
in spite of the rather small number of disorder configurations.

Note also that there is a bias in the determination of the root mean
square current $\Irms$ that is very similar to the bias in $\Upsilon$
discussed above. For $\Irms = \sqrt{\av{I^2}}$ the origin of this bias
is that the statistical error $\delta I$ gives a term $(\delta I)^2
\geq 0$. Since $\delta I$ would be expected to vanish with simulation
time as $\sim 1/\sqrt{\taver}$ the bias in $\Irms$ would vanish as
$1/\taver$, which is essentially the same as the behavior of
$\Upsilon(\taver)$. There is however a difference in that the bias in
$\Irms$ is easily eliminated by performing two independent
simulations, $\alpha$ and $\beta$, for each disorder and measuring the
quantity $\av{I_\alpha I_\beta}$\cite{Olson_Young}. Since the bias in
$\Upsilon$ is of a very different origin it cannot be eliminated with
such methods.

To summarize, we have performed a finite size scaling analysis of the
helicity modulus in a frustrated 3D XY model with disorder in the
coupling constants. The data is consistent with isotropic scaling and
the correlation length exponent is found to be $\nu=1.50\pm 0.12$. The
good agreement with $\nu=1.39\pm0.20$ of the 3D gauge glass suggests
that the two models actually do belong to same universality class.

The author would like to acknowledge helpful discussions with
S.~Teitel and M.~Wallin. This work has been supported by the Swedish
Research Council, contract No.\ E 5106-1643/1999, and by the resources
of the Swedish High Performance Computing Center North (HPC2N).


\begin{thebibliography}{14}
\expandafter\ifx\csname natexlab\endcsname\relax\def\natexlab#1{#1}\fi
\expandafter\ifx\csname bibnamefont\endcsname\relax
  \def\bibnamefont#1{#1}\fi
\expandafter\ifx\csname bibfnamefont\endcsname\relax
  \def\bibfnamefont#1{#1}\fi
\expandafter\ifx\csname citenamefont\endcsname\relax
  \def\citenamefont#1{#1}\fi
\expandafter\ifx\csname url\endcsname\relax
  \def\url#1{\texttt{#1}}\fi
\expandafter\ifx\csname urlprefix\endcsname\relax\def\urlprefix{URL }\fi
\providecommand{\bibinfo}[2]{#2}
\providecommand{\eprint}[2][]{\url{#2}}

\bibitem[{\citenamefont{Fisher}(1989)}]{Fisher:89}
\bibinfo{author}{\bibfnamefont{M.~P.~A.} \bibnamefont{Fisher}},
  \bibinfo{journal}{Phys. Rev. Lett.} \textbf{\bibinfo{volume}{62}},
  \bibinfo{pages}{1415} (\bibinfo{year}{1989}).

\bibitem[{\citenamefont{Fisher et~al.}(1991)\citenamefont{Fisher, Fisher, and
  Huse}}]{Fisher_Fisher_Huse}
\bibinfo{author}{\bibfnamefont{D.~S.} \bibnamefont{Fisher}},
  \bibinfo{author}{\bibfnamefont{M.~P.~A.} \bibnamefont{Fisher}},
  \bibnamefont{and} \bibinfo{author}{\bibfnamefont{D.~A.} \bibnamefont{Huse}},
  \bibinfo{journal}{Phys. Rev. B} \textbf{\bibinfo{volume}{43}},
  \bibinfo{pages}{130} (\bibinfo{year}{1991}).

\bibitem[{\citenamefont{Reichhardt et~al.}(2000)\citenamefont{Reichhardt, van
  Otterlo, and Zim\'anyi}}]{Reichhardt_Otterlo_Zimanyi}
\bibinfo{author}{\bibfnamefont{C.}~\bibnamefont{Reichhardt}},
  \bibinfo{author}{\bibfnamefont{A.}~\bibnamefont{van Otterlo}},
  \bibnamefont{and} \bibinfo{author}{\bibfnamefont{G.~T.}
  \bibnamefont{Zim\'anyi}}, \bibinfo{journal}{Phys. Rev. Lett.}
  \textbf{\bibinfo{volume}{84}}, \bibinfo{pages}{1994} (\bibinfo{year}{2000}).

\bibitem[{\citenamefont{Huse and Seung}(1990)}]{Huse_Seung}
\bibinfo{author}{\bibfnamefont{D.~A.} \bibnamefont{Huse}} \bibnamefont{and}
  \bibinfo{author}{\bibfnamefont{H.~S.} \bibnamefont{Seung}},
  \bibinfo{journal}{Phys. Rev. B} \textbf{\bibinfo{volume}{42}},
  \bibinfo{pages}{1059} (\bibinfo{year}{1990}).

\bibitem[{\citenamefont{Reger et~al.}(1991)\citenamefont{Reger, Tokuyasu,
  Young, and Fisher}}]{Reger_TYF}
\bibinfo{author}{\bibfnamefont{J.~D.} \bibnamefont{Reger}},
  \bibinfo{author}{\bibfnamefont{T.~A.} \bibnamefont{Tokuyasu}},
  \bibinfo{author}{\bibfnamefont{A.~P.} \bibnamefont{Young}}, \bibnamefont{and}
  \bibinfo{author}{\bibfnamefont{M.~P.~A.} \bibnamefont{Fisher}},
  \bibinfo{journal}{Phys. Rev. B} \textbf{\bibinfo{volume}{44}},
  \bibinfo{pages}{7147} (\bibinfo{year}{1991}).

\bibitem[{\citenamefont{Olson and Young}(2000)}]{Olson_Young}
\bibinfo{author}{\bibfnamefont{T.}~\bibnamefont{Olson}} \bibnamefont{and}
  \bibinfo{author}{\bibfnamefont{A.~P.} \bibnamefont{Young}},
  \bibinfo{journal}{Phys. Rev. B} \textbf{\bibinfo{volume}{61}},
  \bibinfo{pages}{12467} (\bibinfo{year}{2000}).

\bibitem[{\citenamefont{Hukushima and Nemoto}(1996)}]{Hukushima_Nemoto}
\bibinfo{author}{\bibfnamefont{K.}~\bibnamefont{Hukushima}} \bibnamefont{and}
  \bibinfo{author}{\bibfnamefont{K.}~\bibnamefont{Nemoto}},
  \bibinfo{journal}{J. Phys.\ Soc.\ Jpn.} \textbf{\bibinfo{volume}{65}},
  \bibinfo{pages}{1604} (\bibinfo{year}{1996}).

\bibitem[{\citenamefont{Kawamura}(2000)}]{Kawamura:00}
\bibinfo{author}{\bibfnamefont{H.}~\bibnamefont{Kawamura}},
  \bibinfo{journal}{J. Phys.\ Soc.\ Jpn.} \textbf{\bibinfo{volume}{69}},
  \bibinfo{pages}{29} (\bibinfo{year}{2000}).

\bibitem[{\citenamefont{Vestergren et~al.}(2002)\citenamefont{Vestergren,
  Lidmar, and Wallin}}]{Vestergren_Lidmar_Wallin}
\bibinfo{author}{\bibfnamefont{A.}~\bibnamefont{Vestergren}},
  \bibinfo{author}{\bibfnamefont{J.}~\bibnamefont{Lidmar}}, \bibnamefont{and}
  \bibinfo{author}{\bibfnamefont{M.}~\bibnamefont{Wallin}},
  \bibinfo{journal}{Phys. Rev. Lett.} \textbf{\bibinfo{volume}{88}},
  \bibinfo{pages}{117004} (\bibinfo{year}{2002}).

\bibitem[{\citenamefont{Olsson and Teitel}(2001)}]{Olsson_Teitel:xy3fp}
\bibinfo{author}{\bibfnamefont{P.}~\bibnamefont{Olsson}} \bibnamefont{and}
  \bibinfo{author}{\bibfnamefont{S.}~\bibnamefont{Teitel}},
  \bibinfo{journal}{Phys. Rev. Lett.} \textbf{\bibinfo{volume}{87}},
  \bibinfo{pages}{137001} (\bibinfo{year}{2001}).

\bibitem[{\citenamefont{Olsson}(1995)}]{Olsson:self-cons.long}
\bibinfo{author}{\bibfnamefont{P.}~\bibnamefont{Olsson}},
  \bibinfo{journal}{Phys. Rev. B} \textbf{\bibinfo{volume}{52}},
  \bibinfo{pages}{4511} (\bibinfo{year}{1995}).

\bibitem[{\citenamefont{Fisher et~al.}(1973)\citenamefont{Fisher, Barber, and
  Jasnow}}]{Fisher_Barber_Jasnow}
\bibinfo{author}{\bibfnamefont{M.~E.} \bibnamefont{Fisher}},
  \bibinfo{author}{\bibfnamefont{M.~N.} \bibnamefont{Barber}},
  \bibnamefont{and} \bibinfo{author}{\bibfnamefont{D.}~\bibnamefont{Jasnow}},
  \bibinfo{journal}{Phys.\ Rev.\ A} \textbf{\bibinfo{volume}{8}},
  \bibinfo{pages}{1111} (\bibinfo{year}{1973}).

\bibitem[{\citenamefont{Li and Teitel}(1989)}]{Li_Teitel:89}
\bibinfo{author}{\bibfnamefont{Y.-H.} \bibnamefont{Li}} \bibnamefont{and}
  \bibinfo{author}{\bibfnamefont{S.}~\bibnamefont{Teitel}},
  \bibinfo{journal}{Phys. Rev. B} \textbf{\bibinfo{volume}{40}},
  \bibinfo{pages}{9122} (\bibinfo{year}{1989}).

\bibitem[{Var()}]{Variance}
\bibinfo{note}{To get an unbiased estimate of the variance of a distribution
  based on $N$ \emph{independent} values $x_i$ one should make use of $\sigma^2
  = 1/(N-1)\sum (x_i -\overline{x})^2$. The prefactor here differs from the
  naively expected $1/N$ since the value of the average $\overline{x}$ in
  general is different from the true average of the distribution. The
  connection with Eq.~(\protect\ref{Ut}) follows since this may be written
  $\sigma^2_\mathrm{naive}(N) = (1-1/N)\sigma^2$.}

\end{thebibliography}

\end{document}